\documentclass[amsmath,twocolumn,showpacs,nofootinbib]{revtex4}   
\usepackage{graphicx}
\usepackage{bm}

\begin{document}

\title{Brownian motion near a liquid-like membrane}
\author{Thomas Bickel}
\email{th.bickel@cpmoh.u-bordeaux1.fr}
\affiliation{CPMOH,
 Universit\'{e} Bordeaux 1 -- CNRS (UMR 5798) \\
351 cours de la Lib\'eration, 33405 Talence, France}

\begin{abstract}

\noindent The dynamics of a tracer molecule
near a fluid membrane is investigated, with particular emphasis  given to 
the interplay between the instantaneous position of the particle
and membrane fluctuations.
It is found that hydrodynamic interactions creates memory effects 
in the diffusion process. 
The random motion of the particle is then shown to cross over from a ``bulk''
to a ``surface'' diffusive mode,
in a way that crucially depends on the elastic properties of the
interface.

\end{abstract}


\maketitle

\section{Introduction}

A largely unsolved problem in soft materials is to understand
how the random motion of a colloid is
correlated with the viscoelastic properties of the embedding fluid~\cite{larsonbook}.
Since complex fluids store elastic energy, the instantaneous position and velocity
of the Brownian particle depend on its prior history.
As a signature of this ``memory'' effect, the friction force experienced
by the particle is, in a generalized Langevin description,
non-local in time~\cite{masonPRL95,amblardPRL96}.
To illustrate this point, we consider a system simple enough
that allows us to extract the memory kernel.
We focus in this paper on the diffusion of nano-particles 
near a liquid-like membrane.
Fluid membranes are soft surfaces, self-assembled from surfactant solutions.
In living systems, they divide the cell into dinstinct compartments
and incorporate a large amount of
proteins involved in signalling activities.
Interactions between nano-sized particles and fluid membranes are 
ubiquitous in phenomena such as antibiotic delivery, 
nano-particle toxicity or virus entry in cells~\cite{biobook}.
This issue is relevant not only from biological or
biophysical standpoints, but also in the formulation of many cosmetics 
and pharmaceutical products~\cite{xieNatmat02}. 
At a more fundamental level, recent experiments on the dynamics of particles
near a membrane have revealed peculiar kinetics effects such as anomalous 
diffusion~\cite{fradinBPJ03}, or
hierarchical transport mechanisms~\cite{kimuraJPCM05}.
Besides, anisotropic coherent motion of particles induced by an alterning electric field
has also been shown to enhance the smectic order in a lamellar phase of 
membrane~\cite{yamamotoNM05}. 
From all these perspectives, a detailed investigation
of the Brownian motion in the vicinity of a membrane appears to 
be essential.

Interfacial contributions to transport coefficients 
have been evaluated long ago for elementary geometries~\cite{happelbook}.  
Far from the surface, the mobility of
a spherical particle of radius $a$ is  $\mu_0=1/(6\pi \eta a)$,
with $\eta$ the shear viscosity of the embedding (simple) fluid.
When a  horizontal wall is located at distance $z_0$ from the particle,
the fluid flow resulting from the motion of the bead 
``reflects'' from the interface
back to the particle, thus exerting an additional
friction force.
At leading order, hydrodynamic corrections 
lead to a position-dependent mobility tensor
\begin{subequations}
\begin{eqnarray}
\mu_{xx}&=&\mu_{yy} =\mu_0 \left[1- \alpha_{\parallel} \left(\frac{a}{z_0}\right)\right]
\ , \label{mux}\\
\mu_{zz} &=&\mu_0 \left[1-\alpha_{\perp} \left(\frac{a}{z_0}\right)\right] \label{muz}  \ ,
\end{eqnarray}
\end{subequations}
and $\mu_{ij}=0$ for $i \neq j$.
Note that within the assumption of point-like particle $a \ll z_0$, 
the exact shape of the colloid is essentially irrelevant~\cite{dhontbook}.
In Eqs.~(\ref{mux}) and (\ref{muz}), particle mobility is affected in a manner
that depends on the nature of the interface.
For a solid surface, the no-slip boundary condition 
 leads to~\cite{happelbook} 
\begin{equation}
\alpha_{\parallel}^{(s)}=\frac{9}{16} \text{ ,  and }
\alpha_{\perp}^{(s)}=\frac{9}{8} \ .
\label{alphas}
\end{equation}
During the last decade, precise experimental measurements of particle 
diffusivity near solid surfaces
have shown excellent agreement with theory~\cite{lanPRL86,faucheuxPRE94,pralleAPA98,dufresnePRL00}.
If, however, the interface separates two liquids having the same 
viscosity $\eta$, the corrections are~\cite{leeJFM79}
\begin{equation}
\alpha_{\parallel}^{(l)}=\frac{3}{32} \text{ ,  and }
\alpha_{\perp}^{(l)}=\frac{15}{16} \ .
\label{alphal}
\end{equation}
More recently, the effects of partial slip~\cite{laugaPF05} as well as fluid 
compressibility~\cite{felderhofJCP05}  have  also  been
included in the calculation of the mobility tensor.

Unlike surfaces having a frozen structure, fluid membranes
are responsive materials. The Brownian motion
of a particle near a membrane is then expected to be dynamically coupled to
the conformations of the interface.
In Section~\ref{hydro}, we consider the problem of free diffusion near a membrane, 
with particular emphasis given to hydrodynamic interactions. 
We show in Section~\ref{mobility} that the instantaneous solution of the 
creeping flow equations
depends on the whole history of both colloid's and interface's motions.
Consequences for the diffusion process
are then discussed in Section~\ref{brownian}.
In particular, we find that the random motion of 
a free colloid crosses over from a ``bulk''
to a ``surface'' diffusive mode. Finally, we come back to the relationship 
with experiments and draw some concluding remarks in Section~\ref{concl},
the technical details being relegated to the appendices.

\section{Hydrodynamics near a fluid membrane}
\label{hydro}

\subsection{Physics of membranes}

Fluid membranes belong to the general class of soft objects (including polymers, gels \ldots),
that deform easily when submitted to external stresses, 
and  undergo thermally 
excited shape fluctuations to increase their configurational entropy.  
At length-scales much larger than the bilayer thickness
(\emph{i.e.}, micrometers \emph{vs.} nanometers), a fluid membrane can be described
as an infinitely thin liquid layer with essentially no in-plane shear viscosity~\cite{seifertEPL93}. 
The instantaneous conformation of the
almost flat membrane is specified through the displacement field 
$h(\boldsymbol{\rho },t)$, with $\boldsymbol{\rho }=(x,y)$. For our purpose, it is
more convenient to use the (2D) Fourier representation 
\begin{equation}
h_{q}(t)=\int d^2 \boldsymbol{\rho } 
\exp[-i\mathbf{q}.\boldsymbol{\rho }] h(\boldsymbol{\rho },t) \ ,
\end{equation}
with $\mathbf{q}=(q_x,q_y)$. 
As first explained by Helfrich, the large variety of shapes and topologies
assume by membranes is governed by bending energy.
For relatively smooth deformations, 
the statistical mechanics of a single membrane
is based on the following Hamiltonian~\cite{seifertAP97} 
\begin{equation}
\mathcal{H}_{m}\left[ h\right] =\frac{\kappa}{2}\int \frac{d^2\boldsymbol{q}}{(2\pi )^{2}}
\left( q^{4} + \xi_{\parallel}^{-4} \right) \left\vert h_{q}\right\vert ^{2}\ ,
\label{helfrich}
\end{equation}
with $\kappa $ the bending rigidity. 
In this description, the membrane is confined in a 
harmonic potential. The in-plane correlation length
$\xi_{\parallel}$ plays the role of an external control parameter setting
the mean square displacement of the membrane:
$ \langle h(\boldsymbol{\rho})^2\rangle^{1/2} \propto
(k_BT/ \kappa )^{1/2}\xi_{\parallel}$, with
$T$  the absolute temperature and $k_B$
the Boltzmann constant. 
As recalled in Appendix~\ref{appA},
the relaxation rate of a fluctuation
with wavevector $\mathbf{q}$ follows the
dispersion relation $\omega_q= \kappa(q^4+\xi_{\parallel}^{-4})/(4 \eta q)$~\cite{zilmanPRL96}.
Overdamped surface waves are further 
characterized by their longest relaxation time
\begin{equation}
\tau_m= \frac{4 \eta}{\kappa} \xi_{\parallel}^3 \ .
\label{deftaum}
\end{equation}
For a typical bilayer in water with $\eta = 10^{-3}$~Pa.s and $\xi_{\parallel}=0.5$~$\mu$m, 
this relaxation time varies between $\tau_m \approx 10^{-3}$~s
for $\kappa = 4.10^{-19}$~J 
and $\tau_m \approx 0.1$~s for $\kappa = 4.10^{-21}$~J.

In order to describe the dynamic organization of a colloidal suspension near a membrane, 
$\tau_m$ has to be compared with the typical time $\tau_d$ needed for a particle to 
diffuse over its own radius
\begin{equation}
\tau_m= \frac{ \eta a^3}{k_BT}  \ .
\label{deftaud}
\end{equation}
This time-scale ranges from 
$\tau_d \approx 10^{-6}$~s for a particle with radius of a few nanometers,
up to $\tau_d \approx 1$~s for micrometer beads. 
Comparison of both time-scales defines two diffusion regimes
in a straightforward manner.
When $\tau_d/\tau_m \gg 1$, \textit{i.e.} when $a \gg \xi_{\parallel} $ or
$\kappa \gg k_BT$, the membrane appears essentially flat to the colloid
and the system is dominated by the relaxation dynamics of the elastic interface. 
We call this regime the weak fluctuation (WF) regime. 
On the other hand when $\tau_d/\tau_m \ll 1$, 
the bead is strongly advected by the 
random flow caused by thermal undulations of the membrane. 
We will refer to the strong fluctuation (SF) regime,
that will be the particular scope of investigation of this paper.

\subsection{Linear hydrodynamic}

For small-amplitude motions, the flow velocity
$\mathbf{u}$ and the local pressure $p$ are governed by the linearized
Navier-Stokes equation
\begin{equation}
\eta  \nabla^2 \mathbf{u} - \boldsymbol{\nabla} p + \mathbf{f} + 
\boldsymbol{\Phi} =\mathbf{0}  \ , 
\label{navier}
\end{equation}
together with the incompressibility condition $\boldsymbol{\nabla} .\mathbf{u}=0 $.
Here, $\mathbf{f} $ is an external force density causing the fluid motion,
and $\boldsymbol{\Phi}$ the restoring force density  
exerted by the deformed interface. The latter
is always perpendicular to the membrane. Within the assumption of smooth deformations,
the force density is given in Fourier space by the functional derivative
$\boldsymbol{\Phi }=(0,0,-\delta \mathcal{H}_{m}/  \delta
h_{q}^{\ast })$.
The creeping flow Eq.~(\ref{navier}) is then solved assuming that
the membrane is not permeable to the fluid, a condition 
quite sensible on experimental time-scales (minutes or hours).
Explicitely, this requirement reads
\begin{equation}
\frac{\partial h}{\partial t} = u_z (\boldsymbol{\rho},h(\boldsymbol{\rho},t),t)  \ .
\end{equation}
The hydrodynamic problem defined in this way is both time-dependent
and highly non-linear due to the fact that $h(\boldsymbol{\rho},t)$
is unknown. Althought it cannot be solved exactly (excepted by numerical methods),
approximate solutions can be found. For instance, iterative methods have proved to be fruitful
when the deformation of the interface is asymptotically small~\cite{yangIJMF90}.
The zeroth-order term usually corresponds to the condition
$u_z (\boldsymbol{\rho},0,t) =0 $, thus representing the motion of
a spherical bead near a flat surface. 
The ensuing velocities and stresses can then be used to
determine a first nonzero approximation for the deformation of the interface~\cite{yangIJMF90}.
This strategy is however limited as it only accounts for the first ``image'' correction
to hydrodynamic interactions.
Instead, we follow a different route and assume that the fluid
motion caused by thermal fluctuations of the membrane is
only slightly perturbed by the presence of the colloid.
The dynamics of the surface  enters the problem through
the  condition
\begin{equation}
\frac{\partial h}{\partial t}   = 
u_z \left(\boldsymbol{\rho},0,t \right) \ .
\label{bc}
\end{equation}
The consistency of this approach is then ensured as long as non-linear
contributions in the deformation field can be neglected.
In fact, this condition has the same range of validity as
the harmonic description for the membrane Hamiltonian~\cite{comment}.

\section{Mobility of a sphere near a fluid membrane}
\label{mobility}

\subsection{Green's function}

Standard Green's function methods are then used to solve  
Eq.~(\ref{navier}). We first evaluate the response function, 
$\mathcal{G}$, to a point-like force
$\mathbf{f}_0 \delta \left( \mathbf{r}-\mathbf{r}' \right)$.
The fluid flow corresponding to a given force field $\mathbf{f}(\mathbf{r})$ is then obtained from 
the convolution
\begin{equation}
\mathbf{u}\left( \mathbf{r},t \right) =  \int dt\int d^3\mathbf{r}' 
\mathcal{G}\left( \mathbf{r},\mathbf{r'}, t-t' \right)
\mathbf{f}\left( \mathbf{r}',t'\right) \ .  
\label{convgreen}
\end{equation}
For the hydrodynamic problem defined above,
the Oseen tensor splits into two contributions, $\mathcal{G}= \mathcal{G}^{(0)} + 
\Delta\mathcal{G}$. The first term, $\mathcal{G}^{(0)}$,
is the well-known free space Green's function
and is given in Appendix~\ref{appA}. The second contribution, $\Delta \mathcal{G}$,
describes the fluid flow caused by the elastic response of the membrane.
While the problem is translationaly invariant in time and in the directions 
parallel to the surface, $\Delta \mathcal{G}$ is expected to depend 
on the perpendicular coordinates $z$ and $z'$ separately.
Incidentally, we remark that the boundary condition Eq.~(\ref{bc}) 
is more conviniently enforced 
if we use the set of variables $ \{ \mathbf{q},z,\omega \} $,
where we define the (time) Fourier transform of any function $F(t)$
as $\widetilde{F}(\omega ) =\int_{-\infty
}^{+\infty }dt e^{-i\omega t} F(t)$.
The main lines of the derivation are given in  
Appendix~\ref{appB}. We find after some algebra
\begin{equation}
\Delta \widetilde{\mathcal{G}}_{kl}(\mathbf{q},z,z',\omega ) 
 = \frac{i}{4\eta q}\frac{\omega_q}{\omega -i \omega_q} \gamma_k(q,z)\gamma_l(q,z') 
\mathcal{M}_{kl} \ ,
\label{deltag}
\end{equation}
where the matrix $\mathcal{M}$
has diagonal elements $\mathcal{M}_{kk}=1$,
and off-diagonal terms $ \mathcal{M}_{xy} =\mathcal{M}_{yx} =1$, 
$\mathcal{M}_{xz}=\mathcal{M}_{yz}=-i$, and 
$\mathcal{M}_{zx}=\mathcal{M}_{zy} =i$.
The functions $\gamma_k$ are given by 
\begin{subequations}
\begin{eqnarray}
\gamma_x(q,z)&=&q_xze^{- q\vert z \vert  }  \ , \label{gx} \\
\gamma_y(q,z)&=&q_yze^{- q\vert z \vert  } \ ,  \label{gy}  \\
\gamma_z(q,z)&=&e^{- q\vert z \vert  }\left(1+ q\vert z \vert \right)  \ .  \label{gz}
\end{eqnarray}
\end{subequations}
Remark that these expressions are consistent with the symmetry property of the 
Green's functions: indeed,  it can be shown from general arguments that
 $\mathcal{G}_{kl}\left( \mathbf{r},\mathbf{r}' \right)=
\mathcal{G}_{lk}\left( \mathbf{r}',\mathbf{r} \right)$~\cite{pozbook}. 
This is re-expressed in terms of our
particular choice of variables as $\widetilde{\mathcal{G}}_{kl}\left( \mathbf{q},z,z',\omega  \right)=
\widetilde{\mathcal{G}}_{lk}\left( -\mathbf{q},z',z,\omega  \right)$,
property that can be checked easily.

In Eq.~(\ref{deltag}), the prefactor $(\omega -i\omega_q)^{-1}$
is a clear signature of hydrodynamic scattering effects.
Indeed, the fluid flow resulting from a displacement of the 
particle exerts stresses that deform the membrane. Relaxing back to its
equilibrium position, the membrane creates a backflow that in 
turn perturbs the motion of the colloid, and so forth. 
The infinite sum $(\omega -i\omega_q)^{-1}=
i/\omega_q\sum_{n=0}^{\infty} (-i)^n(\omega / \omega_q)^n$
is the expression of this infinite series of ``reflexions'' of the 
original point force on the interface.

\subsection{Mobility tensor}

The next step in the derivation consists in identifying the mobility tensor
of a sphere.  
To this aim, we still have to enforce the
no-slip boundary condition for the fluid flow on the surface of the colloid. 
If we note $\mathbf{v}(\mathbf{r}_0)$
and $\boldsymbol{\Omega}$ respectively the translational and rotational velocity 
of the Brownian particle, $\mathbf{r}_0$ being the position of its center-of-mass,
then the fluid velocity is 
\begin{equation}
\mathbf{u}(\mathbf{r}_0+\mathbf{a}) = \mathbf{v}(\mathbf{r}_0) + 
\boldsymbol{\Omega} \times \mathbf{a}   \ ,
\label{clhydro}
\end{equation}
for any vector $\mathbf{a}$ scanning the bead's surface.
The friction force $ \mathbf{F}_H $ exerted by the liquid on the particle
is then obtained from 
\begin{equation}
\mathbf{F}_H = -\int dV \mathbf{f}(\mathbf{r})=-\oint dS \mathbf{f}(\mathbf{a})  \ ,
\label{fhydro}
\end{equation}
where the integral is running over the surface
of the bead. In this expression,
$\mathbf{f}(\mathbf{r})=\mathbf{f}(\mathbf{a})
\delta(\vert \mathbf{r}-\mathbf{r}_0 \vert -a)$ is the force density
exerted by the particle on the fluid. 
Integrating Eq.~(\ref{clhydro}) over the particle's surface,
the angular velocity  cancels out and one obtains,
together with Eq.~(\ref{convgreen}), a linear relation
between the friction force and the velocity of the particle.
This relation defines the (frequency-dependent) mobility tensor,
$\widetilde{\mathbf{v}}(\omega) = - \widetilde{\mu}(\omega)
\widetilde{ \mathbf{F}}_H (\omega) $.
The mobility can be  
written as the sum of two terms,
$\widetilde{\mu}_{kl} (z_0,\omega) = \mu_0 \delta_{kl} + \Delta 
\widetilde{\mu}_{kl} (z_0,\omega)$, with $\mu_0=1/(6\pi \eta a)$ the bulk
value. In the limit of a point-like particle $a \ll z_0$, we obtain 
at leading order
\begin{equation}
\Delta \widetilde{\mu}_{kl} \left( z_0,\omega \right)=\int \frac{d^2\mathbf{q}}{(2\pi)^2}
\Delta \widetilde{\mathcal{G}}_{kl}(\mathbf{q},z_0,z_0,\omega ) \ .
\label{deltamu}
\end{equation}
From  Eqs.~(\ref{gx}) -- (\ref{gz}), one can easily convince oneself
that there are no cross-contributions: $\Delta \widetilde{\mu}_{kl}=0$ if $k \neq l$.

\subsubsection{Static mobility}

The integral Eq.~(\ref{deltamu}) can be performed easily when $\omega=0$.
As can be noticed from Eq.~(\ref{deltag}), the static mobility
does not depend on the elastic properties of the interface.
Indeed, we find
\begin{subequations}
\begin{eqnarray}
\widetilde{\mu}_{xx}(z_0,0)&=&\widetilde{\mu}_{yy}(z_0,0) =\mu_0 \left(1-  \frac{3a}{32z_0}\right)
\ , \label{mumemxx}\\
\widetilde{\mu}_{zz}(z_0,0) &=&\mu_0 \left(1- \frac{15a}{16z_0}\right)  \label{mumemzz} \ ,
\end{eqnarray}
\end{subequations}
We therefore recover the mobility
of a sphere near a flat, liquid interface given in Eq.~(\ref{alphal}).

\subsubsection{Low-frequency limit}

For finite frequency, $\Delta \widetilde{\mu}_{kl}$ is a complicated function of 
the reduced variables $\omega \tau_m$, $a/z_0$ and $\xi_{\parallel}/z_0$.
At this point of the discussion, it is usefull to introduce the friction
coefficient defined as the inverse of the mobility tensor,  
$\widetilde{\zeta}_{kl}=(\widetilde{\mu}^{-1})_{kl}$.
Obviously, the friction tensor is also diagonal.
In the low-frequency limit $\omega \tau_m \ll 1$,
the matrix elements can be expanded and we find, at lowest order,
\begin{equation}
\widetilde{\zeta}_{kl} \left( z_0,\omega \right)=
 \zeta_0 \left[ 1 +\alpha_{kl}^{(l)} \left( \frac{a}{z_0} \right)  
- \beta_{kl} i\omega \tau_m  \right] \ ,
\label{asymp}
\end{equation}
with $\zeta_0=1/\mu_0$.
The prefactors $\beta_{kl}$ depend on the distance to the membrane.
We get $\beta_{\parallel}=0$ and $\beta_{\perp}=3 \pi a/(16 \xi_{\parallel})$
when $\xi_{\parallel} \gg z_0 $,
whereas in the other limit $\xi_{\parallel} \ll z_0$ we obtain  
$\beta_{\parallel}=9a\xi_{\parallel}/(64z_0^2)$
and $\beta_{\perp}=27a\xi_{\parallel}/(32z_0^2)$.

Writing the complex friction as $\widetilde{\zeta}(\omega)
=6\pi \widetilde{\eta}(\omega) a$, it appears that
Eq.~(\ref{asymp}) describes a Maxwell fluid with anisotropic,
frequency-dependent shear viscosity~\cite{larsonbook}. The corresponding relaxation
time is essentially set by the membrane relaxation time $\tau_m$, multiplied by
a geometric prefactor that depends on the distance to the elastic surface.
This relaxation time arises from the delay in the response of the membrane
to a deformation caused by the fluid flow.

\section{Diffusion near a fluid membrane}
\label{brownian}

We are now in a position to study the diffusion of a tracer particle
in the vicinity of a membrane.
Because of hydrodynamic interactions, the friction force experienced by
the Brownian particle is non-local in time.
The corresponding generalized
Langevin equation for a particle of mass $m$ reads
\begin{equation}
m \frac{d \mathbf{v}}{dt} = -\int_{-\infty}^t dt' \zeta (t-t') \mathbf{v}(t') + 
 \mathbf{F}(t)  \ ,
\label{gle}
\end{equation}
with $\mathbf{F}(t)$ the noise term.
All the statistical properties of this random walk can, in principle,
be calculated given the probability distribution of the random force. 
However, the problem is still involved because of the non-linear $z$-dependence
of the friction term. To avoid this complication, we assume that the particle remains
close to its original position during the time of observation  --- the validity of
this assumption being discussed later.
According to the fluctuation-dissipation theorem,
the random force $\mathbf{F}(t)$ can be choosen with zero mean value 
and correlations given by~\cite{kubobook,bedeaux}
\begin{equation}
\left\langle  \widetilde{F}_k(\omega)  \widetilde{F}_l(\omega') \right\rangle
= 2 k_BT \text{Re} \left[ \widetilde{\mu}_{kl} (\omega) \right] 
\times 2\pi \delta (\omega + \omega')  \ .
\label{correl}
\end{equation}

We first focus on the motion in the direction perpendicular to the surface.
In the overdamped limit, \textit{i.e.} for time-scales much longer than $\tau=\zeta_0/m$,
inertial terms can be neglected.  Defining the mean square displacement (MSD) as
$\langle \delta z^2  \rangle = \langle (z(t) -z_0)^2 \rangle$, with $z_0=z(t=0)$, 
we obtain
\begin{align}
\left\langle  \delta z^2 \right\rangle = 2 D_0 t  
- \frac{k_BT}{4 \pi \eta}  \int_0^{\infty}& dq e^{-2qz_0} \left( 1+qz_0\right)^2 \nonumber  \\
&\times \left\{t+ \frac{e^{-\omega_q t}-1}{\omega_q} \right\}  \ .
\label{msd}
\end{align}
Here, $D_0=\mu_0k_BT $ is the bulk diffusion coefficient.
This result presents some interesting features.
At short times, the MSD behaves like 
\begin{equation}
\langle \delta z^2  \rangle \sim 2D_0t \ ,
\label{msdshort}
\end{equation}
for $t \ll \tau_m$. In this limit, the particle does not ``feel'' 
the membrane. On the other hand,
one retrieves for long time-scales the mobility of a particle near a
non-deformable liquid interface
\begin{equation}
\left\langle  \delta z^2 \right\rangle 
\sim
2 D_0\left(1-\frac{15a}{16z_0}\right)  t 
=2 D_{\perp} t   \ .
\label{msdlong}
\end{equation}
for $t \gg \tau_m$.  In this limit, the MSD is thus independent 
of the elastic properties of the surface~\footnote{In fact, 
the long-time diffusion coefficient could also have been obtained from the general relation
$D_{\perp}=k_BT \widetilde{\mu}_{\perp}(\omega=0)$, compare with Eq.~(\ref{mumemzz}).}.
Both the unexpected short-time and the long-time behaviours can however be explained from the definition
of $\tau_m$. As can be noticed from Eq.~(\ref{deftaum}), the limit $t \gg \tau_m$
actually coincides with the limit of an infinitely rigid interface $\kappa \rightarrow \infty$.
Therefore, one ultimately expects to recover the result for a flat interface
when $t \rightarrow \infty$.
On the other hand, the limit $t \ll \tau_m$ corresponds to 
a membrane with vanishing bending rigidity $\kappa \rightarrow 0$.
One can easily check that all the corrections discussed so far actually die out
in that limit, so that we recover the bulk diffusion coefficient when $t \rightarrow 0$.

\begin{figure}
\centering
\includegraphics[width=3.25in]{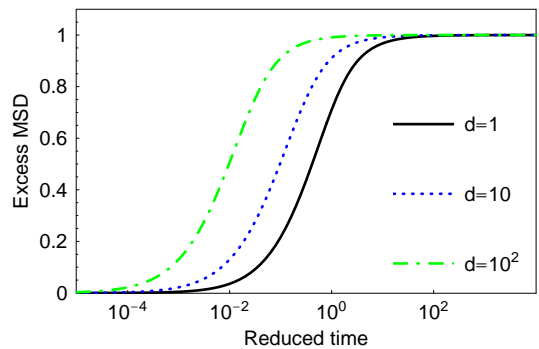}
\caption{Excess mean square displacement $\Delta (t)/ \Delta (t\rightarrow \infty)$ (see text) 
as a function of the reduced time $t/\tau_m$. The curves correspond to different values of the 
reduced distance  $d=z_0 / \xi_{\parallel}$.}
\label{fig2}
\end{figure}

For intermediate time-scales, one crosses over continuously from the fast to the slow diffusion mode.
Defining $\Delta (t)=\langle  \delta z^2 \rangle - 2 D_{\perp}t$, we compute 
numerically and plot in Fig.~\ref{fig2}
the excess MSD $\Delta (t)/ \Delta (t\rightarrow \infty)$ for different values of the ratio
$d=z_0 / \xi_{\parallel}$. When the particle is far away from the membrane, surface deformations
and fluctuations are essentially irrelevant and the particle reaches the ``long-time'' diffusive
regime on a time-scale much shorter than $\tau_m$. However, as the particle approaches the membrane,
the strength of the hydrodynamic coupling between the motion of the particle and
the fluctuations of the membrane increases. The effect of the friction kernel on the Brownian
motion of the colloids can be seen on much longer time-scales, and
the cross-over time increases up to $\sim \tau_m$ for $z_0 \sim \xi_{\parallel}$.
As expected, hydrodynamic effects are all the more relevant as the bending rigidity 
(or the confinement potential) of the 
membrane are small, corresponding to large values of $\tau_m$.

As far as the $x$ and $y$ directions are concerned,
one just has to replace the factor $(1+qz_0)^2$ by $qz_0/2$ in Eq.~(\ref{msd})
to get the corresponding  MSD. 
The results are essentially the same as for the $z$ direction.
In particular, one recovers 
the diffusion coefficient $D_{\parallel}=k_BT \mu_0 (1-3a/(32z_0))$
in the long-time limit
for the same reasons as stated above. 

Finally, we still have to comment on our approximation that 
neglects the $z$-dependence of the friction coefficient. 
This assumption is valid as long as $\langle \delta z \rangle
\ll z_0^2$, that is for $t \ll \tau_0=az_0^2\eta/(k_BT)$.
Therefore, the results described above can be observed provided that
$\tau_0 >\tau_m$, so that we reach the condition
\begin{equation}
\frac{k_BT}{\kappa} <\frac{a z_0^2}{\xi_{\parallel}^3}  \ .
\label{cond}
\end{equation}
For an initial height $z_0=10\xi_{\parallel}$ and 
a typical bending rigidity $\kappa =10k_BT$,
this condition is readily satisfied.

\section{Discussion and conclusion}
\label{concl}

To summarize, we have shown that the random motion of a colloid near
a soft membrane is a non-Markovian process. Indeed, the delay in the 
response of the elastic interface to hydrodynamic stresses
induce memory effects that are relevant over almost three decades in time
(see Fig.~\ref{fig2}). Consequently, 
the Brownian motion of a colloidal particle might locally appear
anomalous with MSD given by $\langle  \delta r^2 \rangle \sim t^{\gamma}$.
But as explained above, hydrodynamic interactions are not expected to 
lead to ``true'' subdiffusive or superdiffusive behaviours since the 
MSD is always linear in time when $t \rightarrow \infty$.

Actually, anomalous diffusion has been observed
in recent fluorescence correlation spectroscopy
experiments close to a vesicle~\cite{fradinBPJ03}.
It has been shown that the intensity autocorrelation function
is the sum of two contributions. 
The short time-scale correlations
arise from the (regular) diffusion of the particles near an impenetrable wall,
whereas the long time-scale correlations come from fluctuations of
the membrane itself. The latter create in turn intensity fluctuations 
by modulating the number of detected particles.
The results are interpreted by considering that diffusion
is \emph{anomalous} due to the collisions of the particles with
the membrane~\cite{fradinBPJ03,granekEPL01}. 
This mechnism is however not incompatible with our results: indeed,
collisions have been neglected so far as they are not easily 
included within a Langevin formalism. In order to get a more
accurate description of the system, both collisions and hydrodynamic 
interactions should therefore be taken into account. 
Note also that the hydrodynamic coupling
described in this work is expected to be most pronounced
for $a \sim z_0 \sim \xi_{\parallel}$. 
A more carefull study, presumably numerical, of these various points 
is therefore required.

In conclusion, we have investigated the motion 
of a Brownian particle in the vicinity of a liquid-like membrane.
The formalism developped in this work allowed us to accurately 
account for membrane-induced hydrodynamic interactions. 
Obviously, this study might be relevant from
a biological perspective
as  diffusive properties of nano-particles near a membrane 
have been shown to crucially depend on the elastic
properties --- or equivalently the composition --- of the bilayer. 
In particular, it would be interesting to see whether
the predicted memory effects can influence the various
kinetic processes taking place near a membrane.

\appendix

\section{Fluctuating hydrodynamics near a fluid membrane}
\label{appA}

Thermal undulations of the membrane actually originate from
hydrodynamic fluctuations of the embedding fluid. 
In this appendix, we re-derive the well-known
equation of motion for the membrane~\cite{zilmanPRL96, seifertAP97} directly from
fluctuating hydrodynamic.
In Fourier representation, the creeping flow equation satisfied by
the flow velocity $\mathbf{u}(\mathbf{k},t)=\int d^3\mathbf{r}e^{-i\mathbf{k.r}}
\mathbf{u}(\mathbf{r},t)$ is~\cite{landau1,landau2} 
\begin{equation}
\mathbf{u}(\mathbf{k},t)=\frac{1}{\eta k^{2}}
\left( \boldsymbol{\mathcal{I}}-\mathbf{\hat{k}}\mathbf{\hat{k}}\right) 
\boldsymbol{\Phi } (\mathbf{k},t) +\mathbf{g}(\mathbf{k},t)\ ,  
\label{stokesfluct}
\end{equation}
with $\boldsymbol{\mathcal{I}}$ the identity tensor, $\mathbf{\hat{k}}
\mathbf{\hat{k}}|_{ij}=\hat{k}_{i}\hat{k}_{j}$ the dyadic product, and 
$\mathbf{\hat{k}}=\mathbf{k}/k$. The restoring force density $\bm{\Phi}=(0,0,\Phi_z)$ is 
given, in Fourier representation,  by
$\widetilde{\Phi}_z(\mathbf{k},\omega) =-\delta H_m/\delta \widetilde{h}_q^*
= -E_q \widetilde{h}_q(\omega)$,
with the energy density 
$E_q=\kappa(q^4 +\xi_{\parallel}^{-4})$
and $\mathbf{q}=(k_x,k_y)$.
In Eq.~(\ref{stokesfluct}), the field $\mathbf{g}(\mathbf{k},t)$ is a
stochastic variable characterizing thermal fluctuations in the liquid.
Following Ref.~\cite{landau2}, it can be shown that the random noise has a
Gaussian distribution with zero mean value, $\langle \mathbf{g}(\mathbf{k}
,t)\rangle =\mathbf{0}$, and correlations 
\begin{equation}
\left\langle g_{i}(\mathbf{k},t)g_{j}(\mathbf{k}^{\prime },t^{\prime
})\right\rangle =\frac{2k_{B}T}{\eta k^{2}}(\delta _{ij}-\hat{k}_{i}\hat{k}
_{j})\delta (t-t^{\prime })(2\pi )^{3}\delta (\mathbf{k}-\mathbf{k}^{\prime})\ ,  
\label{correlations}
\end{equation}
where the square brackets denote average over an equilibrium ensemble.

In order to derive the equation of motion for
the membrane, we still
need to express the impermeability condition
\begin{equation}
\frac{\partial h_{q}}{\partial t}=  u_z\left( \mathbf{q},z=0,t\right)=
 \int_{-\infty}^{\infty} \frac{dk_z}{2\pi}
u_z\left( \mathbf{k},t \right) \ , 
\end{equation}
where the velocity $u_z\left( \mathbf{k},t \right)$ is given in Eq.~(\ref{stokesfluct}). 
Performing the latter integral, 
we obtain the Langevin equation 
describing the relaxation dynamics of a membrane 
\begin{eqnarray}
\frac{\partial h_{q}}{\partial t} -\omega _{q}h_{q}+f_{q}  \ , 
\end{eqnarray}
with $\omega _{q}= E_q/(4\eta q)$. The white noise $f_{q}$ is related
to the stochastic variable $\mathbf{g}(\mathbf{k},t)$ through 
$f_{q}(t)=(2\pi )^{-1}\int $d$k_{z}g_{z}(\mathbf{q},k_{z},t)$. 
Integration of Eq.~(\ref{correlations}) shows that its distribution is still
Gaussian with zero mean value, $\langle f_{q}(t)\rangle =\mathbf{0}$, and
correlations given by 
\begin{equation}
\left\langle f_{q}(t)f_{q^{\prime }}(t^{\prime })\right\rangle =\frac{k_{B}T}{2\eta q}
\delta (t-t^{\prime })(2\pi )^{2}\delta (\mathbf{q}+\mathbf{q}^{\prime })  
\label{memcor}
\end{equation}
Note that these correlations have been obtained after \emph{direct
integration} of the hydrodynamic equations, in the same way as the Langevin
equation of a Brownian particle can be found from fluctuating hydrodynamics~\cite{bedeaux}. 
In particular, the relaxation of a membrane undulation is characterized by 
$\langle |h_{q}(t)|^{2}\rangle =\langle |h_{q}(0)|^{2}\rangle \exp [-\omega _{q}t]$, 
with equilibrium fluctuation amplitude 
\begin{equation}
\left\langle |h_{q}(0)|^{2}\right\rangle =\frac{k_{B}T}{\kappa \left( q^4 + \xi_{\parallel}^{-4}
\right) }\ ,
\label{eqfluct}
\end{equation}
which is expected from the fluctuation-dissipation theorem.

\section{Evaluation of the Green's function}
\label{appB}

In this appendix, we want to derive the correction to the Oseen tensor
coming from hydrodynamic interactions with the membrane.
Because the Navier-Stokes equation is linear, the general solution to an arbitrary force
field $\mathbf{f}$ is given by Eq.~(\ref{convgreen}).
In this problem, the embedding medium is anisotropic so that the
Oseen tensor $\mathcal{G}$ is not translationaly invariant in space.  
The Green function is obtained as the response 
to a point-force applied at point $\mathbf{r}'$ at instant $t$,
$\mathbf{f}(\mathbf{r},t)=\mathbf{f}(t)\delta(\mathbf{r}-\mathbf{r}')$.
After Fourier analysis, the solution of the Navier-Stokes equation together with the 
incompressibility condition reads  
\begin{equation}
\widetilde{\mathbf{u}}\left( \mathbf{k}, \omega \right)
= \frac{1}{\eta k^{2}}
\left( \boldsymbol{\mathcal{I}}-\mathbf{\hat{k}}\mathbf{\hat{k}}\right) 
\left\{ \widetilde{\mathbf{f}}(\omega)e^{-i\mathbf{k}.\mathbf{r}'}+
\widetilde{ \bm{\Phi}} (\mathbf{k},\omega) \right\} \ ,  
\label{stokes}
\end{equation}
with the same notation as in Appendix~\ref{appA}. 
Yet, the deformation field is coupled to the fluid velocity through
the impermeability condition Eq.~(\ref{bc}). This condition is more conviniently 
expressed as $i\omega \widetilde{h}_q(\omega)=\widetilde{u}_z
(\mathbf{q},0,\omega)$, where we used the set of variables $ \{ \mathbf{q},z,\omega \} $.
Finally, coming back to the variables $ \{ \mathbf{k},\omega \} $,
we find the following expression 
for the restoring force density
\begin{equation}
\widetilde{ \Phi}_z \left(\mathbf{k},\omega\right)=
i \frac{E_q}{\omega} \int_{-\infty}^{\infty} \frac{dk_z}{2\pi}
\widetilde{u}_z\left( \mathbf{k},\omega \right) \ .  
\label{force}
\end{equation}
Obviously, this result together with Eq.~(\ref{stokes}) gives
an integral relation for the velocity field. This can be solved
by writing explicitely the equation satisfied by the $z$-component 
of the velocity. Integrating both side of the equation over $k_z$,
we get  
\begin{widetext}
\begin{equation}
\int_{-\infty}^{\infty} \frac{dk_z}{2\pi}
\widetilde{u}_z\left( \mathbf{k},\omega \right) = 
\frac{\exp[-i\mathbf{q}.\bm{\rho}']}{4\eta q} \frac{\omega}{\omega -i\omega_q}
\left[ \left(k_x\widetilde{f}_x+k_y \widetilde{f}_y \right) iz'e^{-q\vert z' \vert}
+\widetilde{f}_z e^{-q\vert z' \vert}\left(1+q \vert z' \vert \right) \right] \ ,  
\label{defi}
\end{equation}
\end{widetext}
with $\omega_q = E_q/(4\eta q)$ and $\bm{\rho}'=(x',y')$.
Finally, bringing together the results Eqs.~(\ref{stokes}),
(\ref{force}) and (\ref{defi}) shows that the velocity
field is proportional to the force, 
$\widetilde{u}_i(\mathbf{k},\omega)=\sum_j \widetilde{\mathcal{G}}_{ij}
(\mathbf{k},\omega)\widetilde{f}_j(\omega)$. The Oseen tensor
can then be written as $\widetilde{\mathcal{G}}=\widetilde{\mathcal{G}}_0
+\Delta \widetilde{\mathcal{G}}$, with the isotropic contribution
$\widetilde{\mathcal{G}}_0 (\mathbf{k},\omega)= 
( \boldsymbol{\mathcal{I}}-\mathbf{\hat{k}}\mathbf{\hat{k}})/ 
(\eta k^{2})$. The explicit expression for the second contribution 
is given in Eq.~(\ref{deltag}) in terms of the variables $\{ \mathbf{q}, z, \omega \}$.


\begin{thebibliography}{99}

\bibitem{larsonbook}  
T. S. Larson, \textit{The Structure and Rheology
of Complex Fluids} (Oxford U. Press, Oxford, 1998).

\bibitem{masonPRL95}  
T. G. Mason and D. A. Weitz, 
Phys. Rev. Lett. \textbf{74}, (1995) 1250.

\bibitem{amblardPRL96}  
F. Amblard, A. C. Maggs, B. Yurke, A. N. Pargellis, and S. Leibler,  
Phys. Rev. Lett. \textbf{77},  (1996) 4470.

\bibitem{biobook}  
H. Lodish, A. Berk, S. L. Zipurski, P. Matsudaira, D. Baltimore, and J. Darnell,
\textit{Molecular Cell Biology}, (Freeman \& Company, New York, 2002).

\bibitem{xieNatmat02}  
A. Feng Xie and S. Granick, 
Nat. Mat. \textbf{1},  (2002) 129.

\bibitem{fradinBPJ03}  
C. Fradin, A. Abu-Arish, R. Granek, and M. Elbaum, 
Biophys. J. \textbf{84}, (2003) 2005.

\bibitem{kimuraJPCM05}  
Y. Kimura, T. Mori, A. Yamamoto, and D. Mizuno, 
J. Phys.: Condens. Matter \textbf{17},  (2005) S2937.

\bibitem{yamamotoNM05}  
J. Yamamoto and H. Tanaka,
Nat. Mat. \textbf{4}, (2005) 75.

\bibitem{happelbook}  
J. Happel, J. and H. Brenner,  
\textit{Low Reynolds Number Hydrodynamics} 
(Kluwer, Dordrecht, 1991).

\bibitem{dhontbook}  
J. K. G. Dhont, 
\textit{An Introduction to 
Dynamics of Colloids} (Elsevier, Amsterdam, 1996).

\bibitem{lanPRL86}  
K. H. Lan, N. Ostrowsky, and D. Sornette, 
Phys. Rev. Lett. \textbf{57},  (1986) 17.

\bibitem{faucheuxPRE94}  
L. P. Faucheux and A. J. Libchaber, 
Phys. Rev. E \textbf{49},  (1994) 5158.

\bibitem{pralleAPA98}  
A. Pralle, E.-L. Florin, E. H. K. Stelzer, and J. H. K. H\"{o}rber, 
Appl. Phys. A \textbf{66},  (1998) S71.

\bibitem{dufresnePRL00}  
E. R. Dufresne, T. M. Squires, M. P.  Brenner, and D. G. Grier, 
Phys. Rev. Lett. \textbf{85},  (2000) 3317.

\bibitem{leeJFM79}  
S. H. Lee, R. S. Chadwick and L. G. Leal, 
J. Fluid Mech. \textbf{93},  (1979) 705.

\bibitem{laugaPF05}  
E. Lauga and T. M. Squires, 
Phys. Fluids \textbf{17},  (2005) 103102.

\bibitem{felderhofJCP05}  
B. U. Felderhof, 
J. Chem. Phys. \textbf{123},  (2005) 184903.

\bibitem{seifertEPL93}  
U. Seifert and S. A. Langer,  
Europhys. Lett. \textbf{23},  (1993) 71.

\bibitem{seifertAP97}  
U. Seifert, 
Adv. Phys. \textbf{46}, (1997) 13. 

\bibitem{zilmanPRL96}  
A. G. Zilman and R. Granek,
Phys. Rev. Lett. \textbf{77},  (1996) 4788. 

\bibitem{yangIJMF90}  
For a review, see S.-M. Yang and L. G. Leal,
Int. J. Multiphase Flow \textbf{16},  (1990) 597.

\bibitem{comment}
It should be noticed that, for the sake of simplicity,
the in-plane incompressibility 
condition for the liquid-like membrane has not been enforced~\cite{seifertAP97}.
Since this condition does not introduce any new 
length- or time-scale, we only expect 
some numerical constants to be possibly affected.
Moreover, it is reasonable to assume that compressibility effects are not
significant in the absence of external shear flow.

\bibitem{pozbook}  
C. Pozrikidis,
\textit{Boundary Integral and Singularity Methods for Linearized
Viscous Flow} (Cambridge U. Press, Cambridge, 1992).

\bibitem{kubobook}  
R. Kubo, M. Toda, and N. Hashitsume,
\textit{Statistical Physics II, Nonequilibrium Statisticl Physics} (Springer Verlag, Berlin, 1985).

\bibitem{bedeaux}  
D. Bedeaux and P. Mazur,
Physica \textbf{76}, (1974) 247.

\bibitem{granekEPL01} 
Incidentally, we remark that anomalous diffusion has also been predicted 
for a particle bound to a fluctuating membrane, see 
R. Granek and J. Klafter,  
Europhys. Lett. \textbf{56}, (2001) 15.

\bibitem{landau1}  
E. M. Lifshitz and L. P. Pitaevskii,
\textit{Statistical Physics Part 2}, Volume 9 of the
\textit{Course of Theoretical Physics} (Pergamon Press, Oxford, 1980), Ch. IX.

\bibitem{landau2}  
L. D. Landau and E. M. Lifshitz,
\textit{Fluid Mechanics}, $2^{nd}$ Edition, Volume 6 of the
\textit{Course of Theoretical Physics} (Pergamon Press, Oxford, 1987).









\end{thebibliography}
\end{document}